\documentclass[sigconf]{acmart}
\AtBeginDocument{%
  }

\setcopyright{acmlicensed}
\copyrightyear{2025}
\acmYear{2025}
\acmDOI{XXXXXXX.XXXXXXX}
\acmConference[AM '25]{Audio Mostly}{June 30--July 04, 2025}{Coimbra, Portugal}
\acmISBN{978-1-4503-XXXX-X/2018/06}

\begin{document}

\title{A case study of translating sonifications across musical cultures for an educational application}

\author{Chris M. Harrison}
\orcid{0000-0001-8618-4223}
\affiliation{%
  \institution{Newcastle University}
  \city{Newcastle}
  \country{United Kingdom}}
  \email{christopher.harrison@newcastle.ac.uk}

\author{James W. Trayford}
\orcid{0000-0003-1530-1634}
\affiliation{%
  \institution{University of Portsmouth}
  \city{Portsmouth}
  \country{United Kingdom}}
\email{james.trayford@port.ac.uk}

\author{Arron George}
\affiliation{%
  \institution{InnerSanctum Entertainment Ltd}
  \city{Point Fortin}
  \country{Trinidad and Tobago}}
\email{innersanctumstudio@gmail.com}

\author{Leigh Harrison}
\affiliation{%
 \institution{Independent}
 \city{Northumberland}
 \country{United Kingdom}}
 \email{leighharr@btinternet.com}

\author{Rub\'{e}n Garc\'{i}a-Benito}
\orcid{0000-0002-7077-308X}
\affiliation{%
  \institution{Instituto de Astrof\'{i}sica de Andaluc\'{i}a}
  \city{Granada}
  \country{Spain}}
\email{rgb@iaa.es}

\author{Shirin Haque}
\orcid{0000-0002-9158-7091}
\affiliation{%
  \institution{University of the West Indies}
  \city{St. Augustine}
  \country{Trinidad and Tobago}}
\email{shirin.haque@sta.uwi.edu}

\author{Rose Shepherd}
\orcid{0009-0002-0369-5146}
\affiliation{%
  \institution{Newcastle University}
  \city{Newcastle}
  \country{United Kingdom}}
\email{r.shepherd12@newcastle.ac.uk}

\renewcommand{\shortauthors}{Harrison et al.}

\begin{abstract}
Sonification can be part of educational resources that can be accessible to those who prefer, or require, non-visual learning methods. Furthermore, sonification can contribute to an engaging multi-sensory learning experience, which are known to benefit general learners. Whilst some sonification can be relatively agnostic to musical culture, many sonifications are subject to culturally influenced choices, such as the chosen harmonies, rhythmic structures, and instrumentation. This is important when considering how universally inclusive and relatable sonification-based educational resources will be. Here we present a case study of translating a sonification-based educational show about the Solar System, that was originally designed with influences from Euro-American (Western-classical) music, to be more culturally relevant to the Caribbean region. We describe the motivation, approach, some of the challenges, and the initial feedback of the resulting output of the project. Finally, we provide reflections on the importance of further work exploring how educational sonifications can transcend international borders and musical cultures. 
\end{abstract}




\received{20 February 2007}
\received[revised]{12 March 2009}
\received[accepted]{5 June 2009}

\maketitle

\section{Introduction}
Representing data with sound, through sonification, as a complement to visually focussed methods has clear potential benefits for education. The most obvious beneficiaries are learners with sight loss; however, there is wider potential for sonification to increase data literacy \citep{Sawe2020}. Indeed, multi-sensory learning enhances educational experiences and memory retention \citep{Shams2008,Okray2023}. 

The benefits of sonification for education are increasingly valued in astronomy and space science; the focus domain of this work. This is evidenced by the increase in astronomy sonification projects over the last two decades \citep{Zanella2022,GarciaBenito2023}. Almost half of these projects’ primary goal is related to education or public communication. Furthermore, many have a primary or secondary goal of increasing accessibility \citep{Zanella2022}. Sonification within astronomy and space science has been internationally acknowledged by the United Nations (UN), through an extensive report on the topic released by the UN Office of Outer Space Affairs \citep{UNOOSA2023}. 

Most astronomy-based sonification projects for education and public communication make use of parameter mapping. This means that a choice needs to be made for mapping data properties to sound properties \citep{Grond2011}. With the intended goals, many sonifications are consequently designed to be aesthetically pleasing and engaging, drawing from musical ideas and techniques, i.e., using data representation methods sometimes named ``musification'' \citep{Barrass2012}. Indeed, as sound is a naturally aesthetic and cultural medium, there is the possibility for sonification to be a popular form of communication, invoking both enjoyment and learning \citep{Barrass2012}. 

The requirement of simultaneous aesthetic appeal and a clear communication of the underlying data can be challenging for the sonification design \citep{Barrass2012,GreshamLancaster2012,Vickers2016}. For example, Grensham-Lancaster (2012) highlights the importance of accounting for personal listening habits and tastes for achieving broader sonification applicability \citep{GreshamLancaster2012}. Indeed, musical perception is deeply influenced by cultural context, including variations in scales, timbres, tuning systems, and even conceptualizations of time. Nonetheless, the role of musical culture in shaping sonification design for educational and communicative purposes has received limited attention. This is a crucial oversight, if sonifications aim to be truly inclusive and accessible to all audiences. Indeed, Garc\`{i}a-Benito and P\'{e}rez-Montero (2022) state the importance of cultural inclusivity in sonification design, emphasizing that many existing artistic sonifications rely predominantly on Euro-American traditions (i.e.,``Western'' aesthetics) \citep{GarciaBenito2022}. As a result, an educational sonification that may be effective for Western audiences, may not feel relatable, relevant, or even understandable, for an audience with different cultural experiences. 

We present a case study of taking an educational sonification-based resource that was tailored for Western audiences and adapting it for audiences in the Caribbean. This was the result of a collaboration between astronomers and musicians based in the UK and in Trinidad and Tobago. The original show can be found here: \url{https://www.youtube.com/watch?v=4jH1WNpDi10}, and our resulting new Caribbean version here: \url{https://www.youtube.com/watch?v=RjkAoqgJvYg}. Downloadable video files are a \url{https://doi.org/10.25405/data.ncl.19345799}, and at \url{https://doi.org/10.25405/data.ncl.29052155}, respectively.

\section{Background}

\subsection{Audio Universe: Tour of the Solar System}
Audio Universe: Tour of the Solar System (AUTSS) is a 35-minute audio-visual educational show tailored to elementary school pupils (aged 7-10) and family audiences. The show was originally released in English and Italian in 2021, and has since been released in Spanish, German, and Portuguese.

The design of the show is described in \citep{Harrison2022} and an evaluation study was conducted \citep{Harrison2023}. Here we give only details necessary to understand the current work. AUTSS was designed in collaboration with members of the sight-loss community, teachers of pupils with sight loss, and the pupils themselves. The soundtrack takes the lead role and is designed to be understandable irrespective of the audiences’ level of vision. The show’s fictional context has the audience riding a spacecraft fitted with a “sonification machine”. This machine turns the light that it detects into sound. Guided by the captain, and a real-life blind astronomer, the audience is taken on an educational audio-visual tour of the Solar System. 

The soundtrack combines sonifications, music, narration, and sound effects. The sonifications communicate key concepts and facts (see Section~\ref{sec:sonifications}). The narration provides storytelling and educational information. The music is primarily used in transitions between scenes when the spaceship is travelling. Both the music and sonifications were created with influences of a Western-classical musical style. This is no surprise, given that the show’s British composer and musical director is a trained classical musician, whose compositions are strongly influenced by those of 20th and 21st century Western-classical composers (\url{https://sites.google.com/view/leighharrison/recordings}). 

AUTSS has reached 10,000s of people (online and in planetaria across multiple countries). An evaluation study of nearly 300 audience members from the UK and Italy revealed that blind audiences celebrated the show’s inclusive nature \citep{Harrison2023}. Importantly, the study also revealed that the sonifications were perceived to be enjoyable and useful by most audiences, irrespective of their level of vision. Nonetheless, these findings cannot be assumed to hold true for audiences with different musical cultures and experiences.

\subsection{Connection to Trinidad and Tobago}
The lead author and AUTSS director (C. Harrison) was invited to Trinidad and Tobago, to present AUTSS and related educational workshops, towards helping provide more inclusive educational resources in the Caribbean region. During this visit, hosted by the University of the West Indies, the AUTSS show’s director met with a local astronomer, a blind musician, physics students, and several people involved in education across Trinidad and Tobago. After experiencing parts of AUTSS, the following reflection was provided by the musician (A. George):

``I overcame countless barriers in pursuing science. There were no facilities or tools designed for someone like me, and I often had to create my own way forward. These sounds opened the Universe to me in a way I could never experience visually, revealing its wonders in a form I could truly appreciate. But I couldn’t help thinking, what if we made this Caribbean-style? Imagine replacing glockenspiels with steel pans, layering vibrant Caribbean rhythms, and infusing our culture into the experience. It would not only reflect the beauty of our region but also make the cosmos accessible and relatable to all who learn differently.''

Consequently, a new collaboration began including with the UK-based AUTSS team, and an astronomer from the West Indies (S. Haque). This resulted in re-designing the soundtrack for AUTSS to make it more relatable for Caribbean audiences, with a focus on the English-speaking West Indies. 

Music is an important part of Caribbean culture, including the world-famous Carnival festivities held in Trinidad and Tobago. Accompanying this is the iconic instrument, the steelpan, developed in the 20th century in Trinidad and Tobago. This region’s music is of great cultural importance, and has given the world reggae from Jamaica, and calypso and soca from Trinidad. Children in Trinidad and Tobago will not be regularly exposed to famous Western-classical musicians, such as Mozart and Beethoven, but will grow up knowing about Montano and Marley. This plays a key role in the philosophical underpinnings of this project: to create a new show with a sense of belonging and relatability to those who have grown up embedded within Caribbean musical culture. 

\section{Methods and Approach}

It was decided to replace the show’s musical soundtrack with music that has a Caribbean feel. The result is a carnival-style musical accompaniment to the travelling spacecraft. This contrasts with the feel of anticipation and drama, invoked by the original Western-classical composition. The ship's captain voice actor was changed from a British voice to a Trinidadian voice, as the accent in the region is very distinct. Finally, as described in more detail below, the sonifications that were identified as the most musical, i.e., those which could be referred to as “musifications”, were replaced.  

\subsection{Sonifications}\label{sec:sonifications}
All the sonifications used in the show were produced using the open-source STRAUSS Python code \citep{Trayford2023,Trayford2025}. STRAUSS is designed to flexibly generate rich data sonifications, through controlling a range of properties of sound. It can use data to directly synthesize audio or to manipulate audio samples. STRAUSS enables flexible spatialization, with all the produced sounds being assigned a directionality (through a polar and azimuthal angle), which are then mapped on to user-defined audio formats, such as stereo and 5.1 surround sound. The sonification designs for the original AUTSS are described in \citep{Harrison2022}; here we describe the changes made for the Caribbean version. 

\begin{figure*}[ht]
  \centering
  \includegraphics[width=\linewidth]{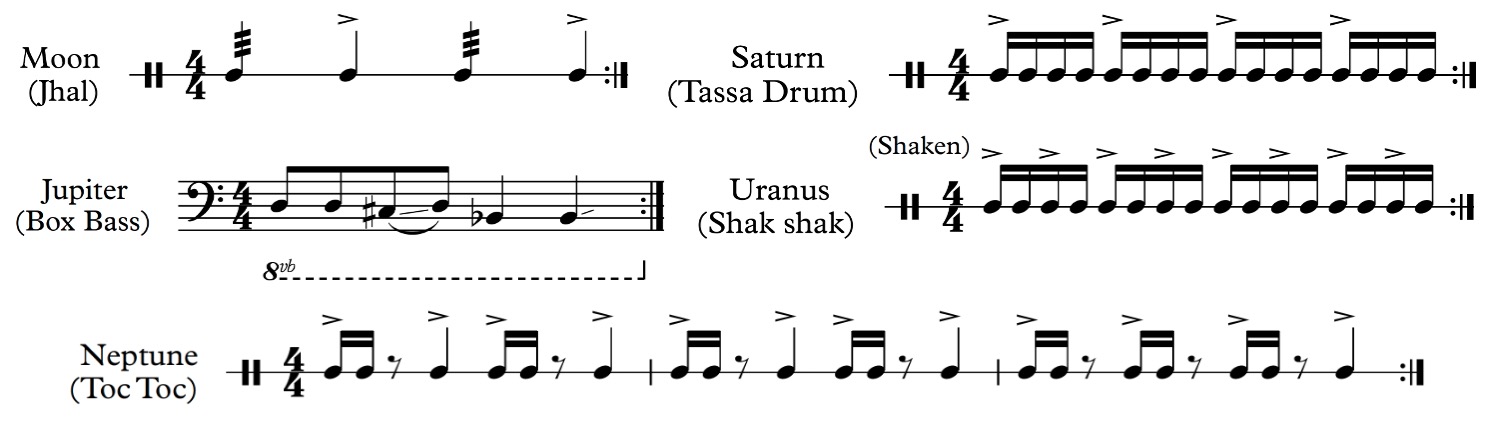}
  \caption{Musical notation of the repeated motifs used to represent Jupiter, Saturn, Uranus, Neptune, and Moon. These are transcriptions of the recordings produced by performers.}
  \Description{Transcriptions of musical motives for each of the Moon, Saturn, Jupiter, Uranus, and Neptune.}
  \label{fig:music}
\end{figure*}

There were three musically-oriented sonifications that were re-designed for the Caribbean AUTSS. All three manipulate audio recordings to communicate the data of interest. In Table~\ref{tab:instruments} and Figure~\ref{fig:music}, we summarize the instruments, musical notes, and rhythms chosen to represent the stars, planets and Moon during these sonifications. For the planets (except Mercury) and the Moon, these are best estimates to notate the actual rhythms and pitches, based on live audio recordings of Trinidadian performers. The sounds to represent the Sun and Mercury were produced entirely electronically. For the stars we used recorded “sound fonts” of single hits on steel pans.

\begin{table*}
  \caption{Instrument choices for astronomical objects in the original AUTSS and the new notes, instruments and rhythms for the Caribbean version.}
  \label{tab:instruments}
  \begin{tabular}{lllll}
    \toprule
    Object & Original Instrument &	New Note(s)	& New Instrument	& New Rhythm\\
    \midrule
    Stars	&Glockenspiel &	[C3,E3,G3,C4,E4,G4,C5]$^{a}$ &	Steel Pan	& Brightness Mapped$^{b}$\\
    Sun (unchanged) &	Synth Bass + SFX & Bb1	& Synth Bass + SFX	& Sustained tone\\
    Mercury &	Flute	& C6 &	Flute&	Sustained tone\\
    Venus	 & Oboe	 &G4	 &Harmonium	 &Sustained tone\\
    Earth &	Clarinet &	F4	 &Steel Pan	 &Roll \\
    Mars &	Soprano Saxophone	 &D5	 &Guitar	 &Trill (Tremolo)\\
    Jupiter &	Bass Trombone	 &B$\flat$1,D2,C$\sharp$2$^{c}$  & Box Bass	 & See Figure~\ref{fig:music}\\
    Saturn	 &Euphonium &	- &	Tassa Drum &	See Figure~\ref{fig:music}\\
    Uranus	 &Trumpet	 &-	 &Shak Shak	 &See Figure~\ref{fig:music}\\
    Neptune	 &French Horn &	- &	Toc-Toc	 &See Figure~\ref{fig:music}\\
    Moon	 &Piccolo &	-	 &Jhal	 &See Figure~\ref{fig:music}\\
  \bottomrule
\end{tabular}
\vspace{0.05cm}\newline
{$^{a}$Chosen note depends on star color (see Section~\ref{sec:sonifications}). \\$^{b}$The effective rhythm for the stars is caused by the parameter mapping (see Section~\ref{sec:sonifications}) 
\\$^{c}$Closest equivalent notes based on actual pitches in the recording. 

}
\end{table*}

\subsubsection{Stars Appearing}
In the show, there is a sequence where the audience experiences the stars appearing after sunset. This sequence appears at times 9m32s--10m58s in the show. The data used for the sonification are the stars’ brightness, colors, and locations on the sky \citep{Harrison2022}. All the stars above a minimum observed brightness (those visible to the eye), are represented by a single note. The actual stars’ sky locations (on a specific time, date, and location) determine the spatial location of the corresponding sounds. The brighter stars are articulated earlier and the dimmer stars later, analogous to the visual experience during sunset, where we see the brightest stars first, and progressively see dimmer stars as the sky becomes darker. Therefore, the rhythm of this sequence is driven by the data itself.  

In the Caribbean AUTTS we use recorded samples from hits on a steel pan to represent each star, in contrast to the glockenspiel samples used in the original. The steel pan is one of the most iconic associations of the Caribbean; therefore, the sequence creates a moment of rich regional associations. We use a C-major triad to select notes. Depending on the star’s actual color, it is mapped to one of 7 notes (all within the triad, but across multiple octaves; Table~\ref{tab:instruments}), with the bluer stars represented by higher notes. The use of a pure major triad, as opposed to the more complex quartal harmony of the original (using notes D$\flat$3, G$\flat$3, A$\flat$3, E$\flat$4 or F4) is more appropriate for the intended musical style. 

\subsubsection{Planet Orbits}
In the original AUTSS, each planet is represented by a sustained note on a unique instrument. The instruments were chosen to distinguish between the four rocky planets (woodwind instruments) and the four gas giants (brass instruments). The notes were chosen to build up a dramatic blend of a B$\flat$ minor chord and a G$\flat$ major chord when played together, along with a synthesized composite sound chosen to represent the Sun \citep{Harrison2022}. 

In the show, the planets are first visited one by one. This is followed by a sequence where all planets are observed together from a distance where their orbits around the Sun can be observed. This sequence occurs during times 27m12s--28m49s in the show. The main information communicated through the sonification is the planets’ relative orbital speeds. The true orbital speeds are scaled up, so that within the 1.5-minute sequence the slowest planet (Neptune) completes a significant fraction of one orbit, whilst the fastest planet, Mercury, can still be tracked in the audio at a reasonable speed \citep{Harrison2022}. The relative orbital speeds are communicated through the sonification by moving the spatial location of the individual sounds in “orbits” around the listener. An additional volume envelope was added to enhance the perceived effect of these orbits, so that the sounds are loudest when directly in front of the listener and quietest when directly behind. 

Choosing familiar instrumentations to represent the planets, with clearly defined notes, was the result of focus group consultations during the design process of the original AUTSS. Audiences preferred to be able to remember and make a direct association between the planets and familiar instruments \citep{Harrison2022}. This helped them maintain the association throughout the show. We carried this philosophy forward for the Caribbean version. However, due to the nature of this project it was important to replace the Western-classical influenced choices with Caribbean-influenced choices. The new instruments chosen to represent each of the planets are listed in Table~\ref{tab:instruments}. 

Whilst the original show made use of purely sustained tones to represent each planet, there was a shift to make more use of percussive instruments for the Caribbean version. We used extended audio recordings to capture the playing style of Trinidadian musicians on local instruments. These were manipulated in intensity, and spatialization, using the STRAUSS code, following the exact same mapping approaches from the original AUTSS. The rocky planets were chosen to be represented by sustained tones, trills, or rolls. The gas giants (which have much longer orbits) are represented by repeated rhythmic motifs, all played at 120 beats per minute (Figure~\ref{fig:music}). The effect is that when all the planets are heard together, there is a dynamic mixture of tones and rhythms, with a distinctly Caribbean feel.

\subsubsection{The Moon}
Similarly to the planets, the Moon is given a distinct characteristic sound. In the original AUTSS a sustained note on a piccolo was used. In the Caribbean version, a Jhal was chosen, with a simple rhythmic pattern (Figure~\ref{fig:music}). At one point in the show, the Moon’s orbit is described, and a sequence is made where the Moon is heard to orbit the Earth. As with the planets, this effect is achieved with spatialization of the sound around the listener in addition to a volume envelope. This Moon orbit sequence occurs during times 16m45s--17m28s.

\section{Challenges}

A few challenges were identified during this project. Firstly, these related to the available facilities in the region. Specifically, whilst the original AUTSS main format was for planetaria, with full-dome visuals and 5.1 surround sound, this is not applicable for the Caribbean region. The complete lack of a planetarium in the West Indies, and the minimal access to surround sound facilities, resulted in quickly shifting the focus to a flat-screen version using a stereo soundtrack. Spatialisation of sound can be achieved in STRAUSS using any speaker set-up \citep{Trayford2023,Trayford2025}; and the stereo effect can still be immersive and effective when wearing headphones. The front-to-back spatialization cannot be directly achieved; however, we applied additional volume envelopes such that objects at the “front” appear louder \citep{Harrison2022}. 

The biggest musical challenge was the important emphasis on rhythms in Caribbean music, which conflicted with the original focus on pitch and timbre to distinguish between the astronomical objects in the sonifications. Indeed, much of the design of the sonifications and musical soundtrack of AUTSS was built around constructing a dramatic chord when the Sun and eight planets are heard together \citep{Harrison2022}. The Sun acts as a musical pedal point, with a continuous drone of a low note. Each of the planets are then assigned a different note, through a continuous tone, which builds upon this pedal point to construct a complex musical chord. These continuous tones can easily be shifted in spatial location, and tracked by the listener, to give the effect of the planets, or Moon, orbiting (Section~\ref{sec:sonifications}). However, we deemed it necessary to use more percussive instruments and rhythmic patterns, as they are so central to Caribbean music. To create an effective sonification sequence of the planets (and Moon) orbiting, put tight constraints on these rhythms. They needed to be continuously tracked by the listener without creating undesirable spatial aliasing when the planet moved significantly in any silent periods between hits. This challenge was not initially anticipated and resulted in multiple iterations between the Trinidadian musicians and the AUTSS team before reaching a workable solution. 

\section{Premiere and Initial Feedback}
It is necessary to understand the educational environment in the English-speaking Caribbean schooling system to put the importance of this project in context for engaging local audiences with astronomy. We note that this region is often lumped together as the Latin America and Caribbean countries (LAC), yet there is a divide of language barrier that isolates the English-speaking Caribbean from the wider LAC \citep{Schonberg2022}. In this region, astronomy does not appear in the syllabi of secondary school education and the primary school (ages 5 – 12) has limited astronomy in its syllabus. The first real educational exposure to astronomy can happen only at university level. Even so, this is almost exclusively at the University of the West Indies in Trinidad and Tobago, where the sole professional astronomer in the English-speaking Caribbean has been based, being joined only recently by another in the Bahamas. However, enthusiastic amateur astronomers have taken up the challenge of filling this vacuum of educational exposure to astronomy, with the Trinidad and Tobago Astronomical Society and the Caribbean Institute of Astronomy (CARINA, co-founded by Shirin Haque) with many outreach events that are well supported. Astronomy societies and clubs are limited to Jamaica, Barbados and Trinidad with many islands being underserved \citep{Haque2006}. Furthermore, outreach for blind and low vision persons has been non-existent until very recently when Shirin Haque and collaborators formed the Caribbean Astronomy for Inclusion group with the support of blind astronomer Dr. Wanda Diaz-Merced in 2021 \citep{OcanaFlaquer2022}. Despite our best efforts, it was still challenging to break through the wall to include these audiences in astronomy events \citep{Haque2022}. It is with this backdrop, there were high levels of enthusiasm and excitement at the launch of the Caribbean AUTSS. 

The Caribbean AUTSS premiered at the University of the West Indies, Trinidad and Tobago, on the 30th of January 2025. This was presented to a diverse audience of approximately 30 people, including students and staff, and representatives working in wider educational and political roles in Trinidad and Tobago. The audience members included those with a range of levels of vision, including those who are fully sighted and fully blind. Approximately 25\% represented the blind and low vision community, a record level for local outreach events. Although a formal evaluation was not conducted, qualitative feedback was very positive. For example, one person noted:
``To see science, sound, music and inclusivity come together in a project that speaks to us in the Caribbean was such a unique and uplifting experience. I look forward to all the avenues yet to be explored using this process.'' 

Motivation was high to build on this project, and make it more widely embedded. For example, one person stated: ``I'd like to see this have its own place in our society, the same way we can say let's go watch a movie on a weekend, we should be able to say let's go have a listen to the audio universe and experience the universe. I see it having great potential for education and the non-academically inclined as well since the overall mood it sets is something not exclusive to learning but something that can be enjoyed leisurely as well.''

Since experiencing the premiere, the Trinidad and Tobago Blind Welfare Association has begun conversations with the authors on how to build on this methodology to support their community in secondary and tertiary education. We expect that the full impact of this project, and future work building upon it, will be seen in the months and years to come.

\section{Final Reflections}
We embarked on a unique project of “translating” a sonification-based educational show, which had significant Western-classical influenced content, to make it more relatable and relevant for Caribbean audiences. The criteria were to produce something that was both “musically satisfying” and “data driven”, where the aesthetics and effective communication were intertwined in the design \citep{Barrass2012}. This naturally led to an iterative design process with multiple interactions between the scientists and musicians involved. This resulted in using an appropriate harmonic palette and putting more emphasis on rhythmic patterns and percussive instruments (compared to the original AUTSSS), reflecting the musical traditions of the Caribbean.

Whilst we have presented some early reflections and qualitative feedback, this work is not intended to provide rigorous proof of success, nor address all the possible challenges and issues related to the topic of the relevance of musical culture in sonification-based educational resources. Instead, we have presented this as an initial proof-of-concept idea and highlighted some of the benefits and challenges we discovered. We hope it encourages others to perform future work on this topic, including a systematic study on the impact of accounting for different musical cultures in the effectiveness of educational sonifications. This could include investigating relatability and perspectives, even within the same culture, such as across-generations or across-genders.

We have highlighted the importance of considering musical culture for increasing the reach and accessibility of sonification designed for education and public communications. By incorporating multiple cultural perspectives, sonifications can create shared experiences that enhance empathy, challenge social divisions, and ultimately reach a wider, more diverse audience \citep{Vuoskoski2017,Gustafson2020}. Garc\`{i}a-Benito and P\'{e}rez-Montero (2022) argue that achieving more inclusive sonifications requires incorporating diverse musical traditions and aesthetic frameworks \citep{GarciaBenito2022}. They emphasize that involving creators from different cultural backgrounds can help ensure that sonifications are expressed through multiple sonic languages rather than being restricted to a single aesthetic paradigm. This helps address the previously discussed idea, that a listener’s background (“schematic knowledge”) might affect a listener’s interpretation of sonification \citep{Roddy2016}. While historical cross-cultural musical interactions have been well documented \citep[e.g.,][]{GarciaBenito2024} and re-enacted in modern performances and recordings \citep[e.g.,][]{GarciaBenito2021}, this approach has only seen limited application in sonification \citep[e.g.,][]{Yu2019,Buehler2020}. This is something we hope will change moving forward. 

Transcending musical cultures with sonification will not be without its challenges. One approach could be to stick to sonification approaches that are less musically driven. For example, associating data with the sounds of physical objects (sometimes called “iconification”; \citep{Barrass2012}), may be one popular alternative \citep{Barrass2010}. However, rigorous evaluations on the effectiveness and audience preferences, with culturally diverse participants, are starkly lacking. 

Barrass (2012) stated that a careful design perspective may lead to a future where the public tunes into “pop sonifications” for both enjoyment as well as obtaining useful information about the world \citep{Barrass2012}.  We share this dream, adding that we must ensure it can transcend international borders and musical cultures.

\begin{acks}
 The authors acknowledge funding from UKRI and STFC (codes: MR/V022830/1, MR/Z506448/1, ST/X004651/1, ST/W006790/1, and UKRI133), the Severo Ochoa grant CEX2021-001131-S (funded by MCIN/AEI/10.13039/501100011033) and grant PID2022-141755NB-I00. We thank the Simons Foundation and Newcastle University for supporting the collaborative trip that initiated this project. For the purpose of open access, the authors have applied a Creative Commons Attribution (CC BY) license to any Author Accepted Manuscript version arising.
\end{acks}

\bibliographystyle{ACM-Reference-Format}

\end{document}